# A STUDY ON THE INTERACTIVE "HOPSCOTCH" GAME FOR THE CHILDREN USING COMPUTER MUSIC TECHNIQUES

Shing-Kwei Tzeng[1] and Chih-Fang Huang [2]

[1]Department of Information Communications, Kainan University, Taiwan
sktzeng@mail.knu.edu.tw
[2] Department of Information Communication, Yuan Ze University, Taiwan
jeffh@saturn.yzu.edu.tw

## ABSTRACT

*"Hopscotch" is a world-wide game for children to play since the times in the ancient Roman Empire and China. Here we present a study mainly focused on the research and discussion of the application on the children's well-know edutainment via the physical interactive design to provide the sensing of the times for the conventional hopscotch, which is a new type of experiment for the technology aided edutainment. The innovated hopscotch music game involves the sound samples of various animals and the characters of cartoon, and the algorithmic composition via the development of the music technology based interactive game, to gradually make the children perceive the world of digits, sound, and music. It can guide the growing children's personality and character from disorder into clarity. Furthermore, the traditional teaching materials can be improved via the implementation of the electrical sensing devices, electrical I/O module, and the computer music program Max/MSP, to integrate the interactive computer music with the interactive and immersive soundscapes composition, and the teaching tool with educational gaming is completely accomplished eventually.*

## KEYWORDS

*Hopscotch, Algorithmic Composition, Edutainment, Interactive Design, Soundscapes*

## 1. INTRODUCTION

The children game "hopscotch" is one of the most well-known traditional games played by the children all around the world including many countries, although the content slightly differs from the various versions derived by different countries or areas [1]. It can be used as an auxiliary tool for children's education purpose to attract the eyes to concentrate their learning. However most of the teaching materials used for the children education are presented in the form with common static methods, lack of the innovative games integrated with technology. In this research a new method to adopt the conventional children game with new music technology is presented, in order to enhance the interactivity for the children education with the innovative computer game aided. The adopted hopscotch game with music technology integrated can be used for the children under the age of ten, with simple electronic sensing system equipped, to inspire the collaborated music composition from sports, arts, and crafts. Bloom and Hanych [2] indicate that equating learning with fun suggests that if students are not enjoying themselves, they are not learning. Therefore the hybrid word "Edutainment" is created to emphasize the game-like formats in the media [3]. Some researchers emphasize that educational technology is a medium, not a pedagogy that is useful in creating learning environments [4]. Some traditional games with new added technology have been developed for the rural children [5], which show the potential of using technology to integrate with the traditional game. More recent research developments in computer music during the last decades are conducting toward more sophisticated and novel musical applications to integrate for intelligent user interfaces, games, entertainment, and edutainment [6].

The purpose to present the idea of adding music technology with the traditional game will not only extend the usage of the traditional hopscotch, but also enhance the learning effect and potential of the aided game for the children education. Many research articles indicate that the education material with music technology aided tools can improve children's concentration and learning efficiency [7, 8, 9, 10]. A successful case of music technology aided game tool for children education is "Hyperscore" [11, 12] presented by MIT Media Lab, which helps children to compose music with the simple curves and graphics easily. In the recent research, computer music aided tools for education has been widely adopted, such as the tool developed by Janet McDowall with MIDI sequencer as the interactive computer music tutorial to reduce the complexity of the music composition for the children [13]. Sensor integration plays a very significant role in the fields of interactive music technology, via the sensing systems including IR, touch control, and RF, etc., to control the input signal from the music game [14]. The human-machine creativity also plays an important role for the revolutionary interactive music composition with proper digital media applied, to develop a new artistic style in the recent years. In addition the various aesthetic discussions among art and technology integration [15] shows the possibilities to form the new art styles with the interactive media, with HMI (Human Man Interface), sensors, and actuators well designed. However from the compute r science point of view, the interactive performing or game already uses the techniques from the interactive data mining [16, 17]. The "Hopscotch" children game with music technology integrated will use computers to synthesize a series of consecutive actions into a kid of collaborated music composition and performing. The powerful computing capability of the innovative games can also improve the education content based on their analysis and the interactivity features [18]. In addition to the formal training of the music composition and performance, some auxiliary skills including the technology aided content including music notation, emotion communication, improvisation, and the limbs exercise can greatly reduce the performers' psychological disorder [19]. Therefore the game with music technology tools can effectively help the education, particularly for the training in the early childhood.

The presented new hopscotch game also adopts the concept of soundscapes, which has been widely used for the electronic music field since the ear of early forties [20, 21]. Soundscapes not only collects the sound from natural environment and machines, but also integrate the pre-recorded tape sound with lots of transformation to compose the music organically. In the study of the new hopscotch game, we preserve the sound samples from cartoon characters and animals into the memory of a personal computer, and design the user interface with touch sensing unit to allow the children to jump with their feet to select the sound samples while their playing. Therefore the multitrack montage soundscapes [22] can be created via the innovated game played interactively, to inspire the imagination from the electronic music automatically. The automated sieve algorithm used for the new hopscotch music game were explored by Iannis Xenakis' "Sieve" theory [23], which was also implemented in Phil Winsor's automated computer music program [24]. Max/MSP developed by Miller Puckette and David Zicarelli, etc., can help the new hopscotch game with the automated music composition [25] and the sensing equipment integration easily, with its high-level graphical control programming method.

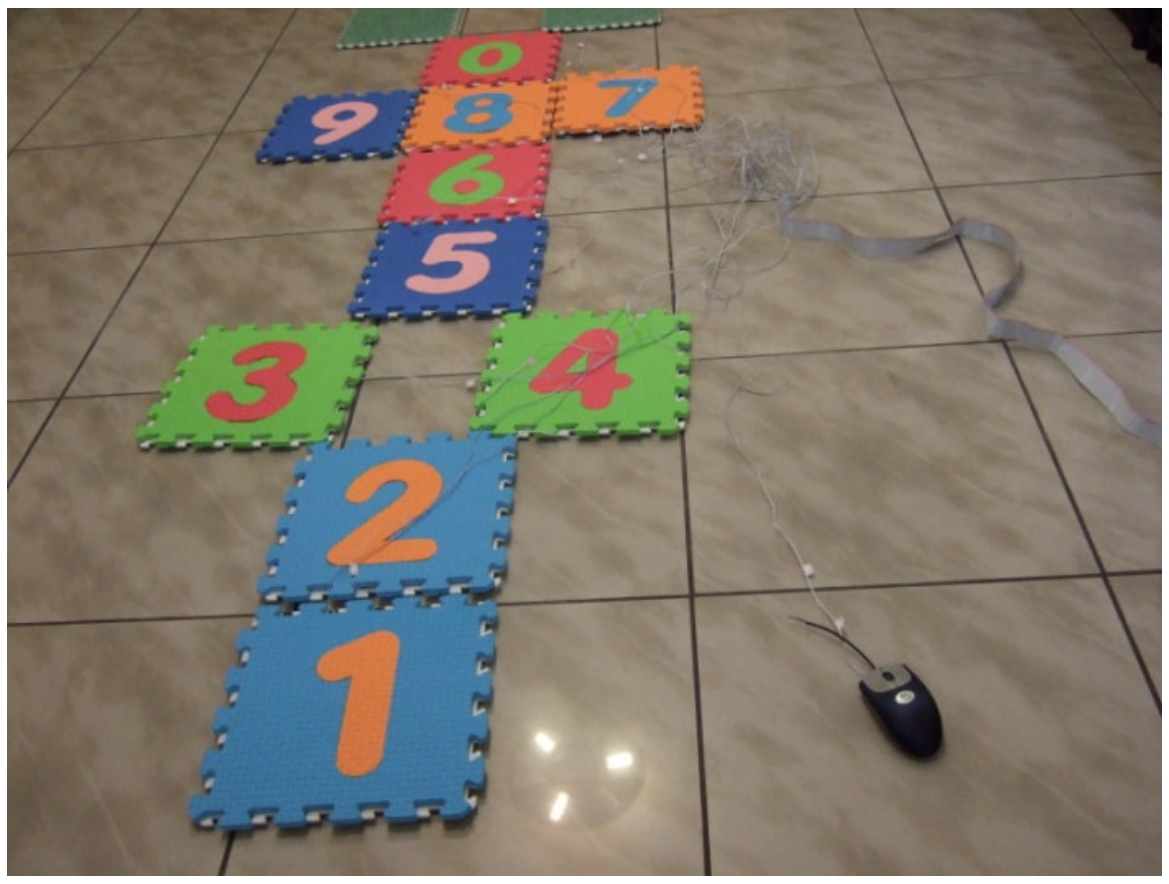

Figure 1. The Innovative Hopscotch Game with Music Technology Techniques.

## 2. METHOD

The new way to implement an innovative hopscotch game with edutainment, music technology, and interactive multimedia techniques are based on the following crucial concepts:

1. New HMI (Human Main Interface): There are twelve touch control switches installed under the digit pad of the game, to provide the trigger of the each sound sample while children play, to compose an immersive and interactive soundscapes music composition. There sound modes can be selected by the PC mouse, including the sound from the cartoon characters, the animals, and the computer generative.
2. Interactivity Promotion: The touch control sensors and the sound system played with the game can greatly increase the interactivity, and the generated sound can be considered as the collaborated composition by the children, which also extend the usage and endow the new life of the traditional hopscotch game.
3. Hyperinstrument Application: The hyperinstrument concept implicitly exists in the new developed hopscotch game, to establish the automated computer music composition with sensor integration, and the interaction between the children and the game.

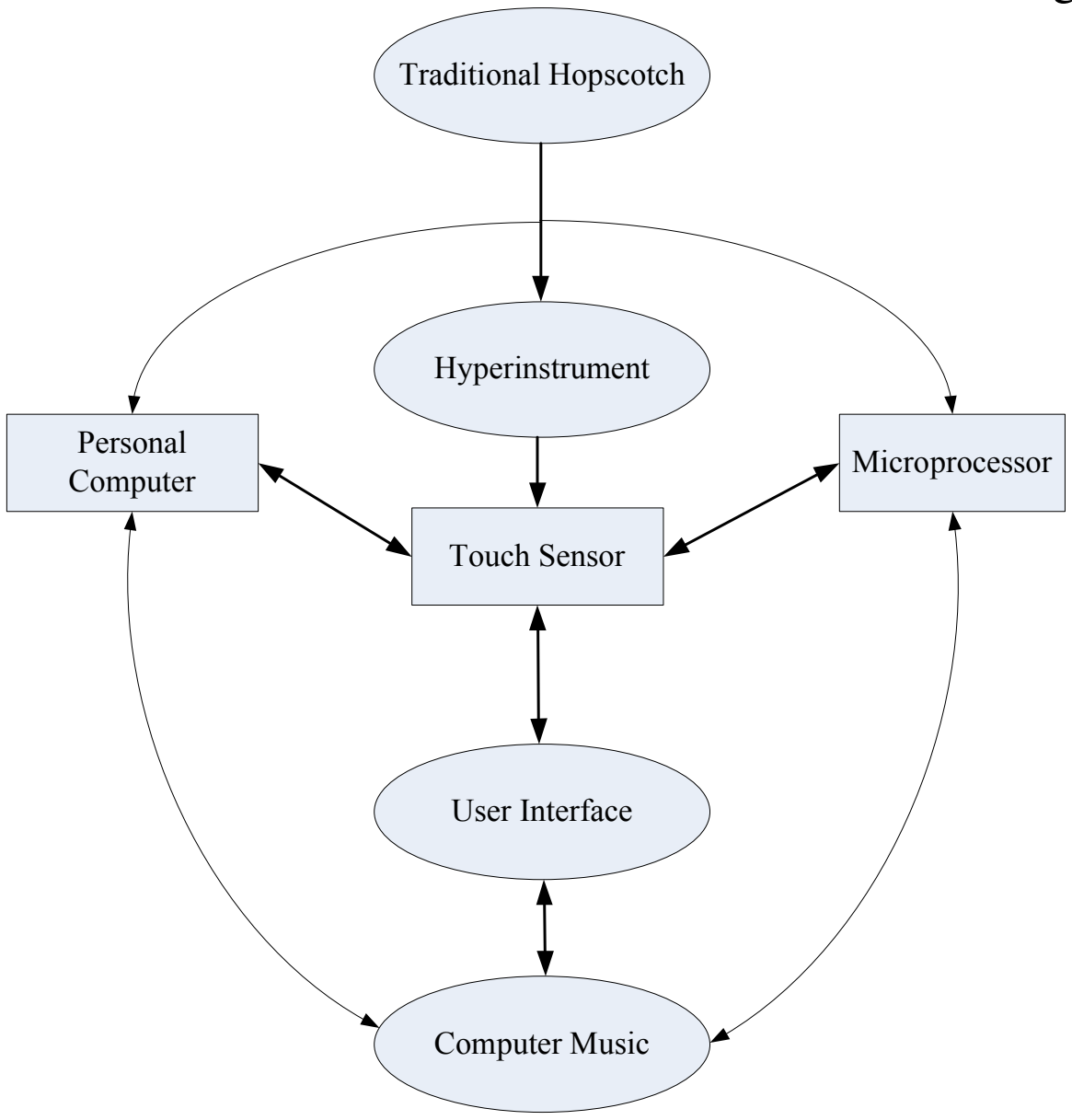

Figure 2. The Main Concept to Establish the New Hopscotch Music Game

The new hopscotch game system architecture mainly includes the following units:

1. PC with Max/MSP Program: the Max/MSP music programming can integrate the electronic module to provide three soundscapes modes with mouse clicked, and detect the jumping touch sensing, and send the sound wave file output to the sound system.
2. Microprocessor with C Language Program: the Microprocessor provide the HMI sensing capability to accept children's jumping position into the touch control switches, and provide the signals to PC for the sound integration.
3. Sound System: the sound samples include cartoon character sound, computer generative sound, and animal sound, which can be selected by the right clicks of the mouse.
4. Digit Pad with Touch Control Sensors: the twelve touch sensing switches are attached beneath the digit pads waiting for children to jump onto them, and then the computer music is generated while the game is played. The touch switches is designed based on the electrical buttons.

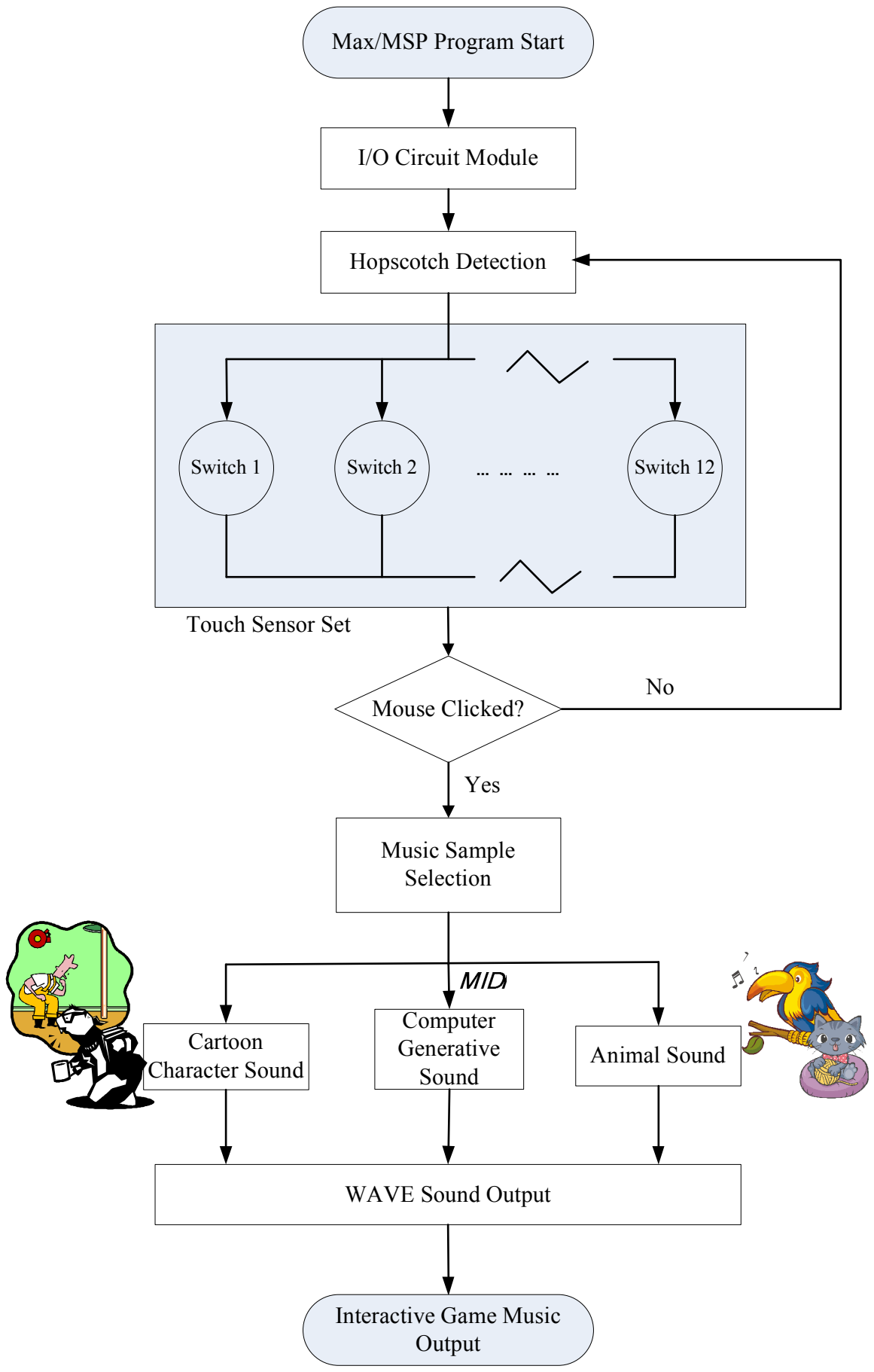

Figure 3. The System Architecture of the New Hopscotch Music Game

## 3. ARCHITECTURE: HARDWARE

The hardware architecture provides a real time mechatronic system for the innovative hopscotch game including three modules as follows:

1. PC module: the art technology based Max/MSP program is resident in the PC module as the main development tool for the computer music, to connect the user's interface module with USB serial port and the sound system, to construct a fully integrated system.
2. UI module: UI (User's Interface) module includes a microprocessor with C language compiled and loaded, to program the 16-bit data bus to communicate with the remote user via the I/O unit to integrate with the PC module, and Sensor module.

3. Sensor module: there are 12 touch sensor switches for each hopscotch digital pad's input, to wait for children's stepping on their feet as the trigger signals with the 16-bit digital bus connected.

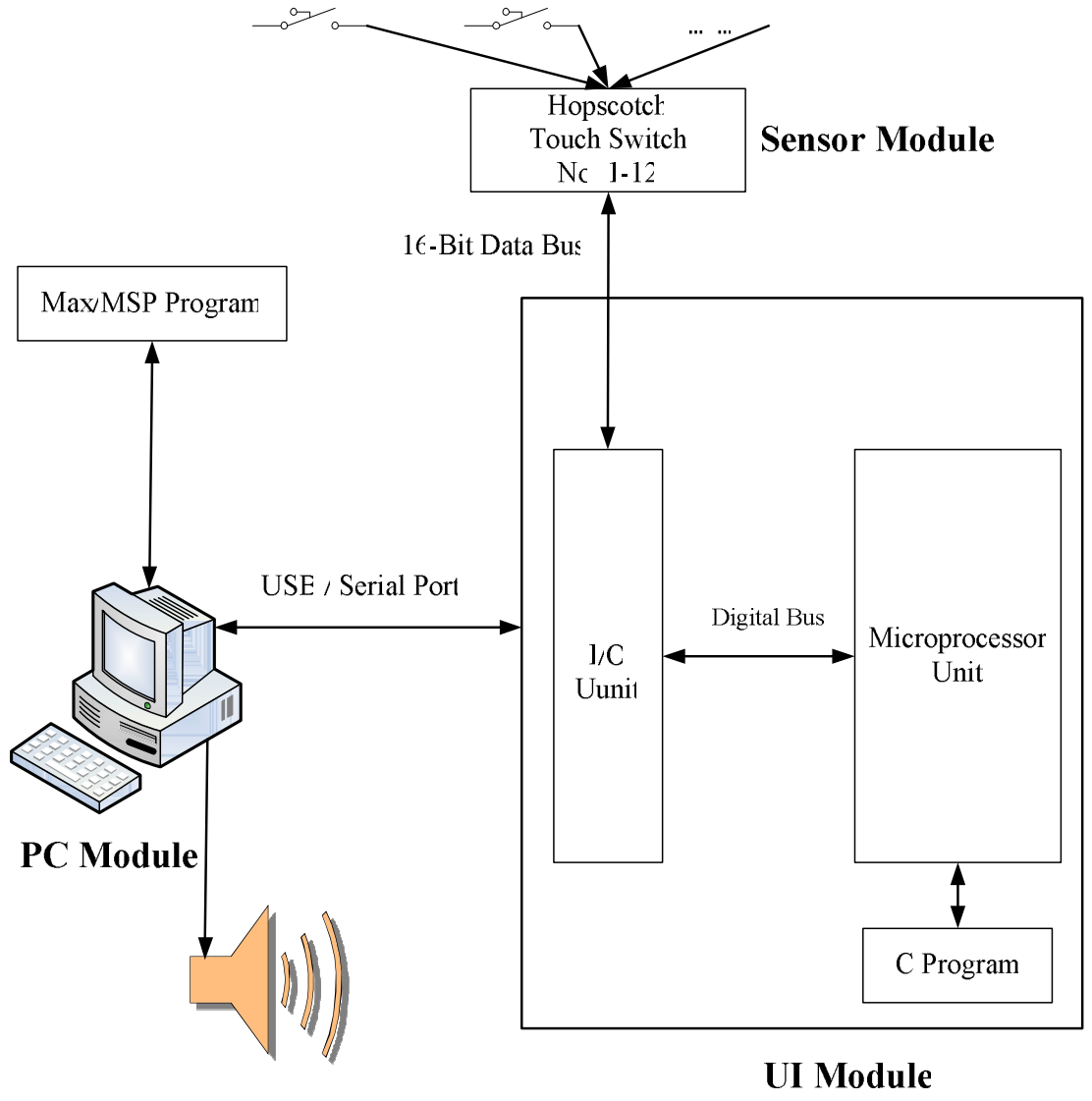

Figure 4. The Three Modules for the New Hopscotch Game Integration

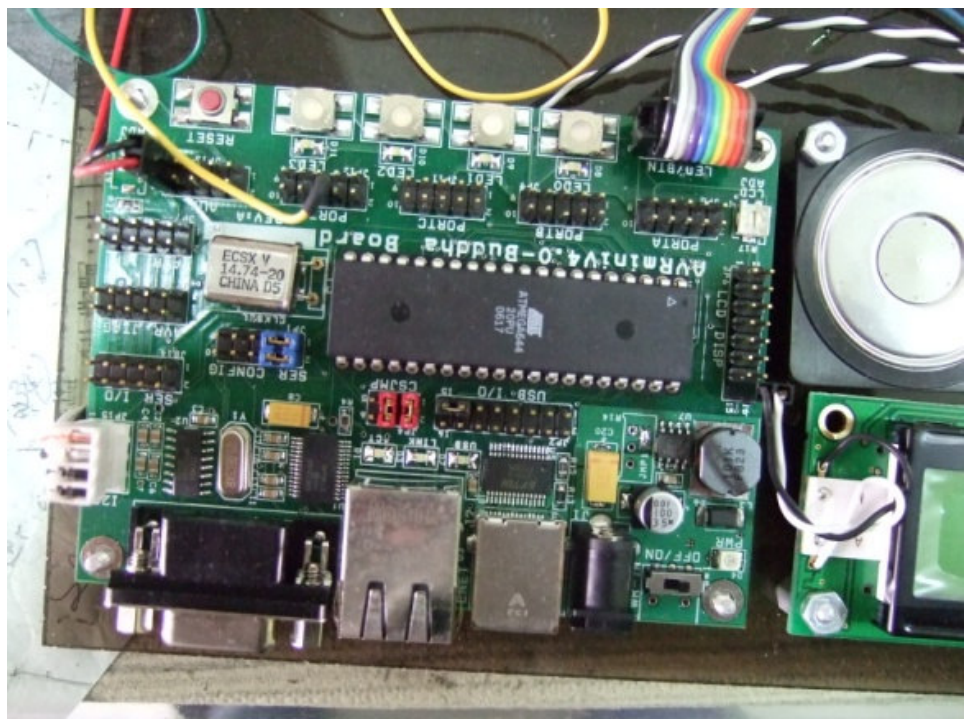

Figure 5. The Microprocessor Unit for the New Hopscotch Music Game

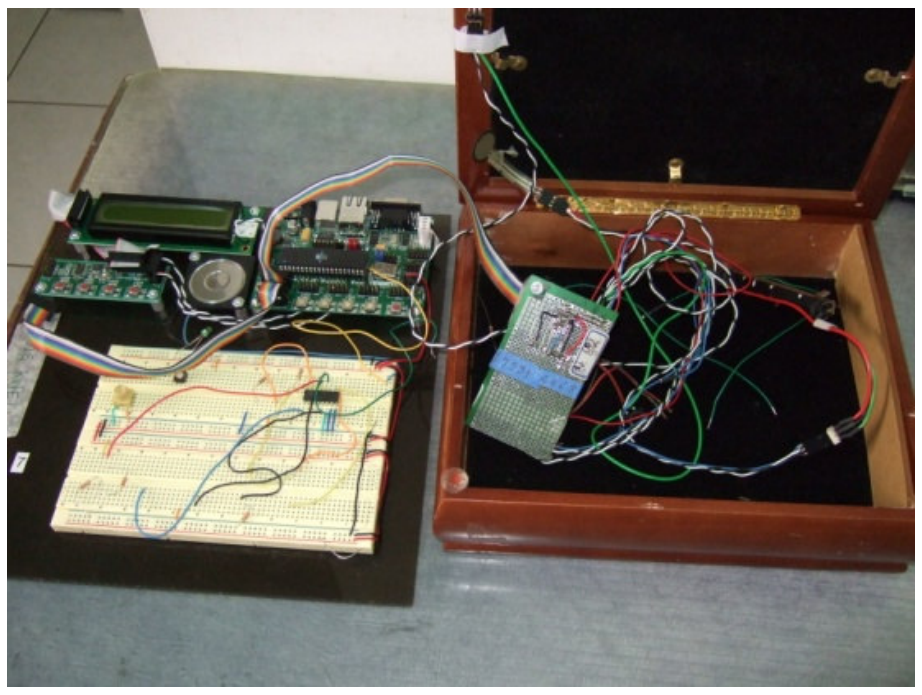

Figure 6. The UI Module for the New Hopscotch Music Game

## 4. Architecture: Software

As to the software implementation, there are two set of programs including the Max/MSP computer music program resident in the PC, and the C language to drive the electronic board with microprocessor loaded.

### 4.1. Max/MSP Program

The Max/MSP program not only performs the music composition, but also implements the data communication function with both UI module and Sensor module. The following figure shows the Max/MSP patch for the I/O mapping for the UI module, to obtain the logic state reported by the touch sensing device.

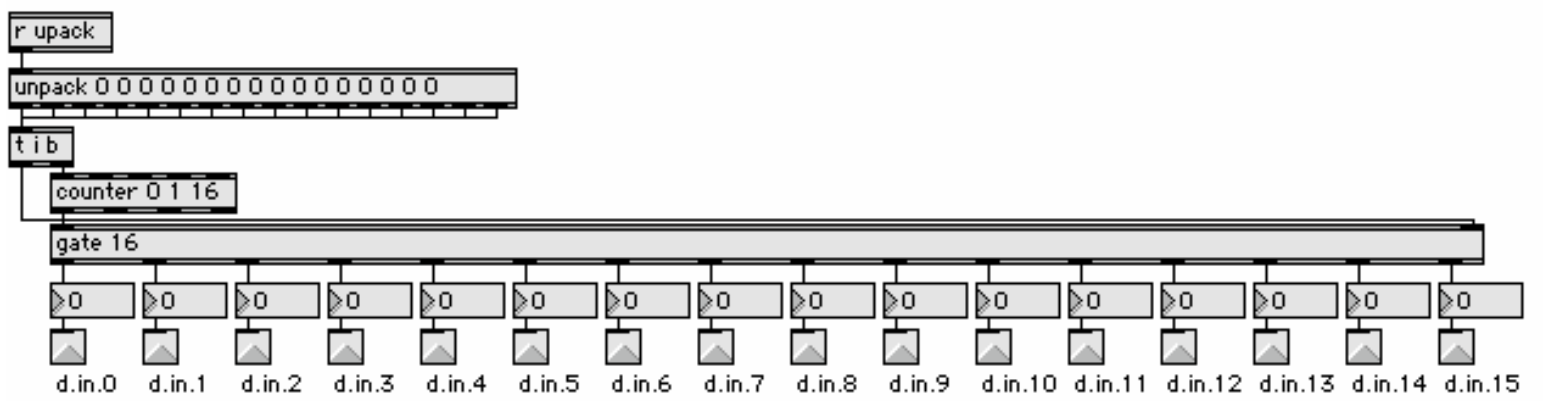

Figure 7. Max/MSP Patch for I/O Mapping

After the I/O mapping is set, the serial port can be initiated with proper baud rate set, for instance 38400Hz, and each touch sensor state can be retrieved via the following Max/MSP patch programmed, to be controlled and integrated for the automated music composition.

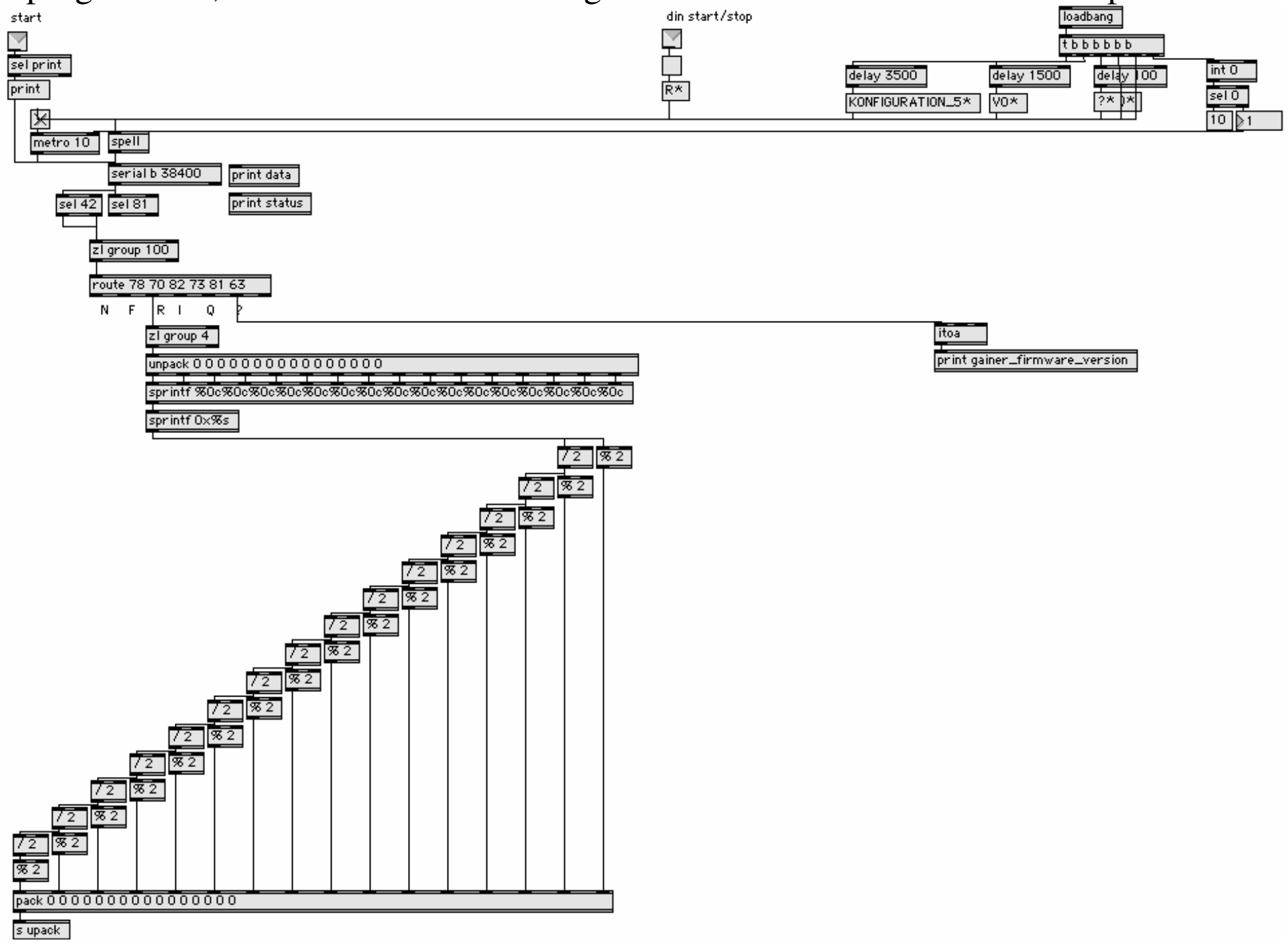

Figure 8. Max/MSP Patch for the Interface Control

The Xenakis Sieve is used to control the computer generative sound for the new hopscotch music game as the automated music algorithm. As shown in Figure 9, modulus 12 is selected to program the music scale, to send the “noteout” MIDI device. If the mouse click for the sound mode select is “Computer Generative Sound”, then the automated music composition will be enabled, and the programmed music scale will be played via the touch sensing from the UI module.

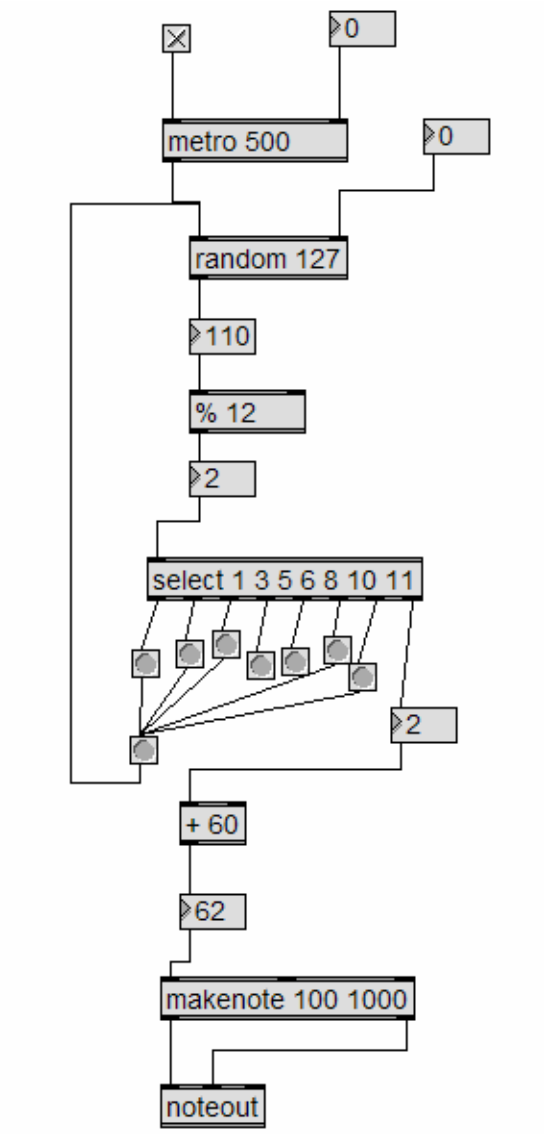

Figure 9. Max/MSP Patch for the Automated Sieve Program

The following Max/MSP patch indicates the computer music sound control status, to acquire each touch sensing state from the UI module and the sound mode selection from the mouse click. The pre-recorded sound samples of the cartoon characters (sounds from subway, drums, button, etc.) or the animal sound (sounds from bird chirp, monkey, and buffalo etc.) will be automatically played accordingly.

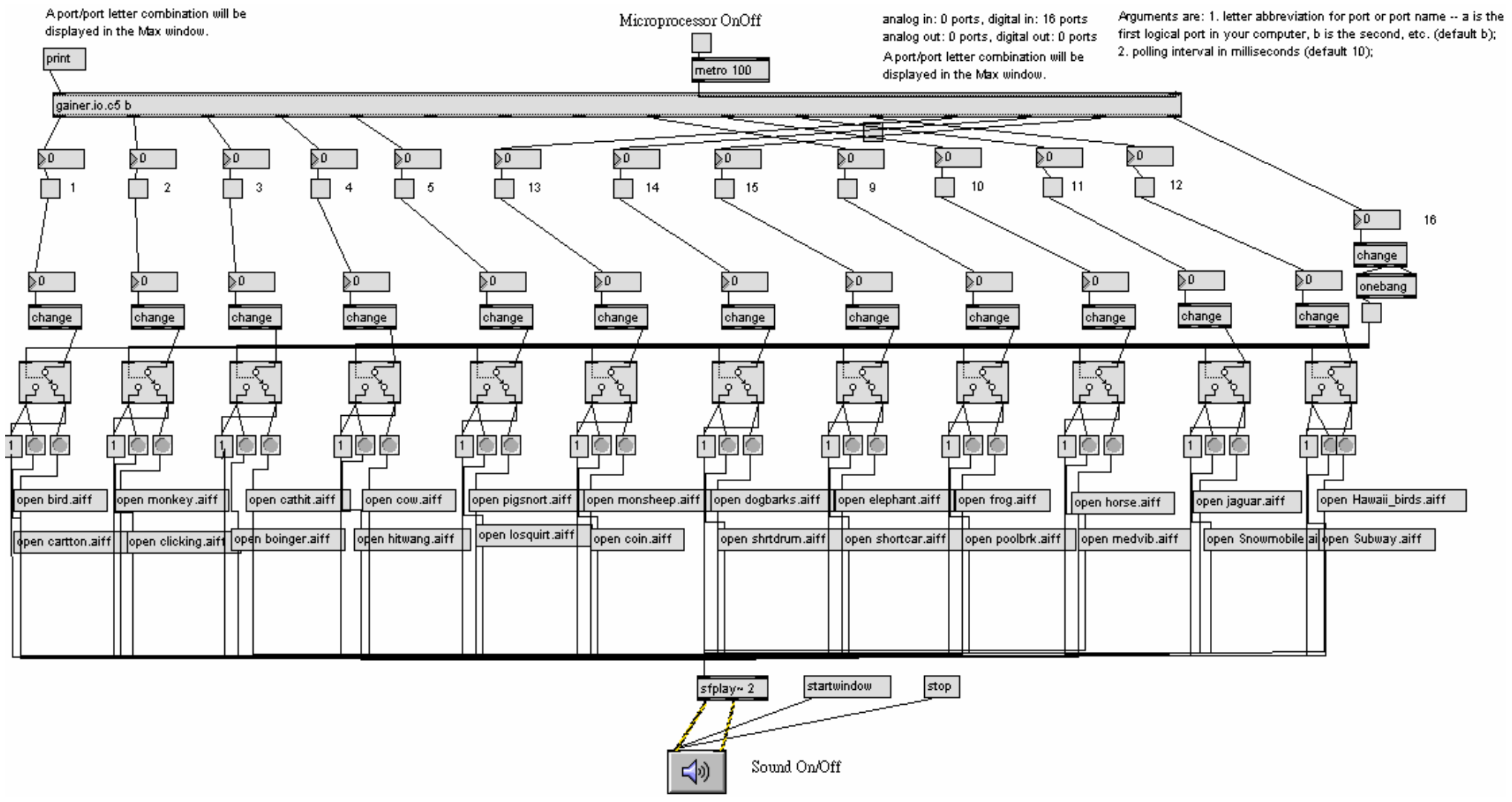

Figure 10. Max/MSP Patch for the Immersive and Interactive Soundscapes Composition

## 4.2. C Language Program in Microprocessor

The logic states of the touch sensing buttons can be triggered by the children via UI module, and sent to the microprocessor with the definition of the touch control logic by the program. The C language program segment is depicted as below:

```
//global variables
u16 Slider_value;
```

```
u16 piezo_value;
u16 bend_value1;
u16 bend_value2;
u16 fsr_value;
u16 optic_value;
#define BUTTON_PORT PINB //port B
#define DEBOUNCE_THRESHOLD 100
u08 buttonPressed( u08 pin ) { // which pin on the port
  if ( !bit_is_set( BUTTON_PORT, pin ) ) {
//button is pressed?
    long debounce = 0;
    while(!bit_is_set( BUTTON_PORT, pin ))
//wait for release
      debounce++;
    if (debounce > DEBOUNCE_THRESHOLD)
      return 1;
  }
  return 0;
}
```

The main procedure of the program can provide the functions of UART serial port communication, oscillator interrupt and timer, to perform the real time detect of the touch control state, and send the data to the PC module. The I/O definition for the data bus and the data format for the communication are also specified in the program. The follows is the code segment:

```
int main(void)
{
  // OSC stuff
  // message flag - when this gets set to one in a function
  //     the main will send a message to OSC
  uartInit();
  oscInit();
  timerInit();
  // LCD STUFF
  /* string to print*/
  //  lcdInit();
  //  rprintfInit(&lcdDataWrite);
  //  lcdClear();
  //  lcdHome();
  // rprintfProgStr(str_up);
  // ATOD STUFF
  a2dInit();
  a2dSetReference(0x01);
  cbi(DDRA, DDA0);
  cbi(DDRA, DDA1);
  cbi(DDRA, DDA2);
  cbi(DDRA, DDA3);
  cbi(DDRA, DDA4);
  cbi(DDRA, DDA5);
  cbi(DDRA, DDA6);
  cbi(BUTTON_PORT, 0);
   //clear bit on the port (set to ground 0V)
  sei();                    /* enable interrupts */
  while (1)  {              /* loop forever */
    timerPause(50);    /* pause to slow down A/D */
     //   cbi(DDRA, DDA1);
    bend_value1 = a2dConvert10bit(4);
    /* set PortA PIN 1 to input */
    bend_value2 = a2dConvert10bit(5);
    fsr_value = a2dConvert10bit(1);
    optic_value = a2dConvert10bit(6);
    piezo_value = a2dConvert10bit(3);
    Slider_value = a2dConvert10bit(2);
     //   sbi(DDRA, DDA1);        /* read ATOD value */
```

```
oscSendMessageOneArg(PSTR("/bend_data1"), bend_value1);
oscSendMessageOneArg(PSTR("/bend_data2"), bend_value2);
oscSendMessageOneArg(PSTR("/optic_data"),optic_value);
oscSendMessageOneArg(PSTR("/piezo_data"),piezo_value);
oscSendMessageOneArg(PSTR("/fsr_data"),fsr_value);
oscSendMessageOneArg(PSTR("/Slider_data"),Slider_value);
   oscSendMessageOneArg(PSTR("/trigger1"), buttonPressed(0));
oscSendMessageOneArg(PSTR("/trigger2"), buttonPressed(1));
 }
  a2dOff();
  return 0;
}
```

With all of the programs and hardware modules integrated, the automated music will be generated while the game is played, and the sound file can be recorded and edited with multitrack audio program, to compose the interesting computer music.

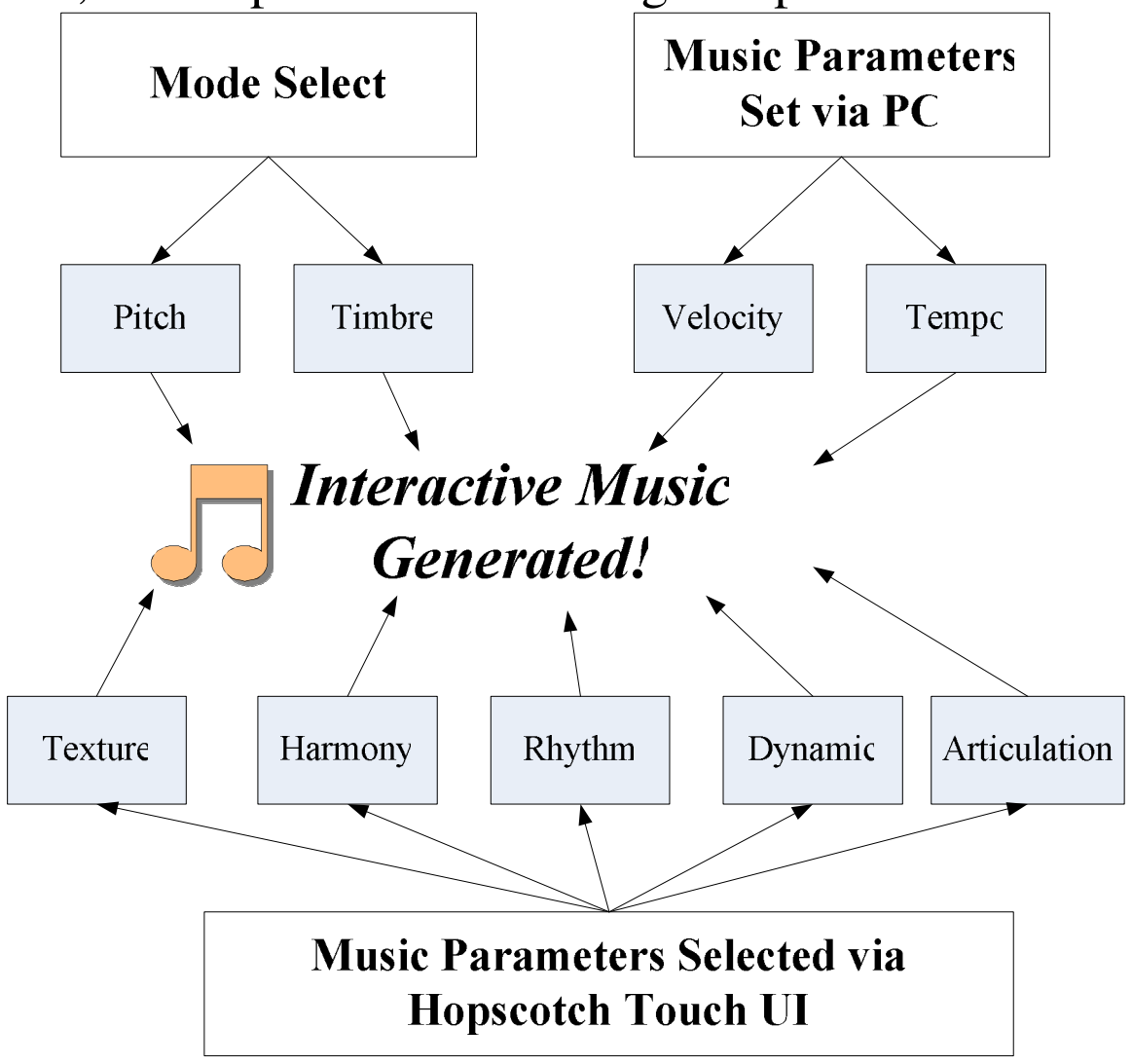

Figure 11. The Automated Computer Music Generated While the Game is Played

## 5. EVALUATION

The key music parameters of the new hopscotch include pitch, rhythm, texture, sound response, dynamic, timbre change, etc. With the supervision of the instructor, all the related music parameter can be adjusted via the response from the children all the time. The game and teaching can be applied to the activities of the group recreation, and the optimized music parameters can be found out based on the recorded data and the children playing response.

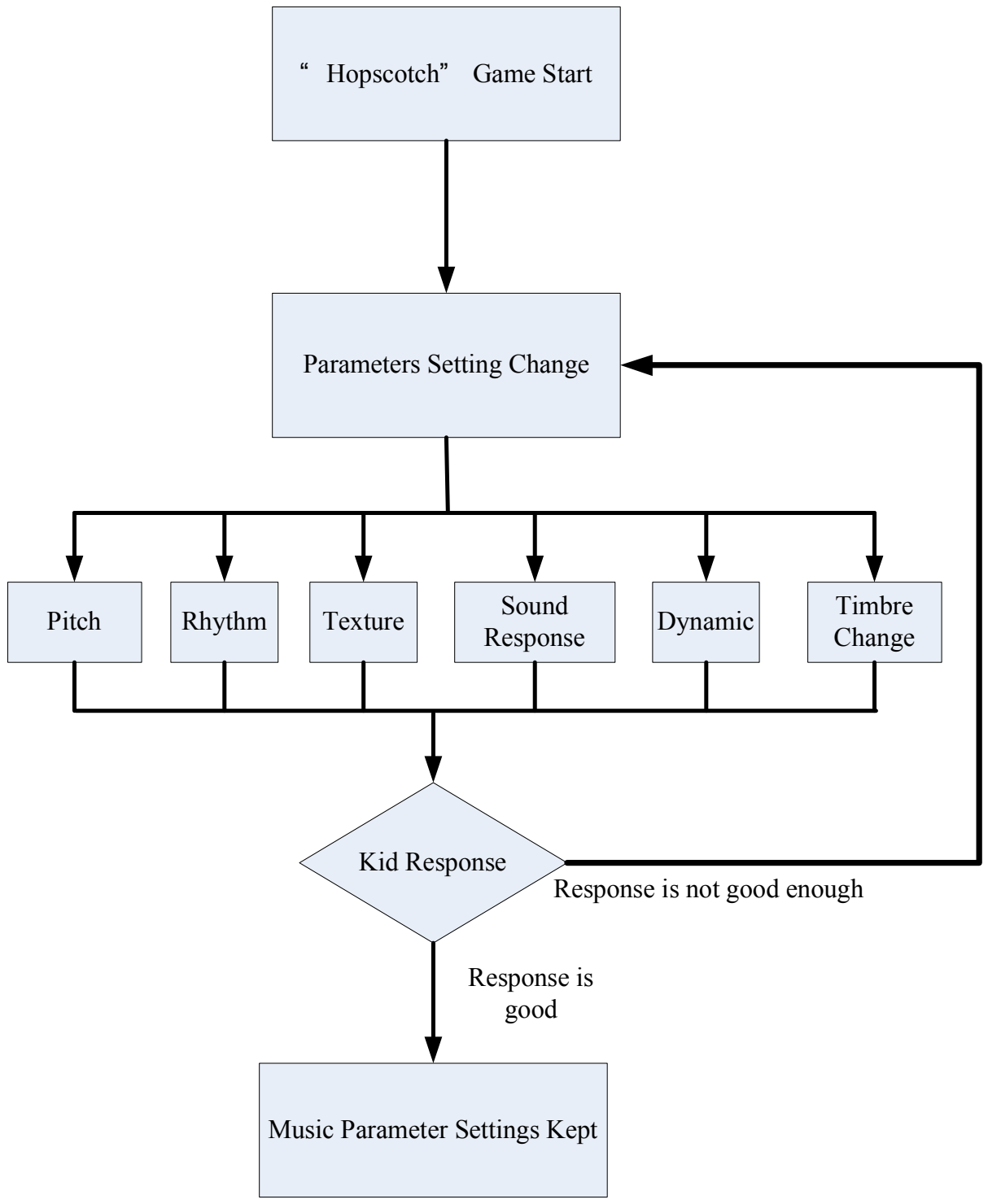

Figure 12. The Music Parameters Settings and Adjustment for the New Edutainment Game "Hopscotch"

According to the general observation by several teachers, the children response for the new hopscotch game is evaluated as the following performance table:

| | Rhythm Variety | Pitch Sensation | Texture Change | Sound Response | Dynamic Variance | Timbre Change |
|---|---|---|---|---|---|---|
| A | ● | | | | | |
| B | | | | ● | ● | ● |
| C | | ● | | | | |
| D | | | ● | | | |
| E | | | | | | |

Note: A – Excellent (90% and above)
B – Good (80%~90%)
C – Medium (70%~80%)
D – Not bad (60%~70%)
E – No good (60% and below)

Table 1. The Performance Evaluation for the Music Parameters of the New Hopscotch Music Game

Based on the above evaluation, the music game can be improved by the polyphonic texture, with the variation of the pitch structure, to reinforce the system performance of the computer music in the future.

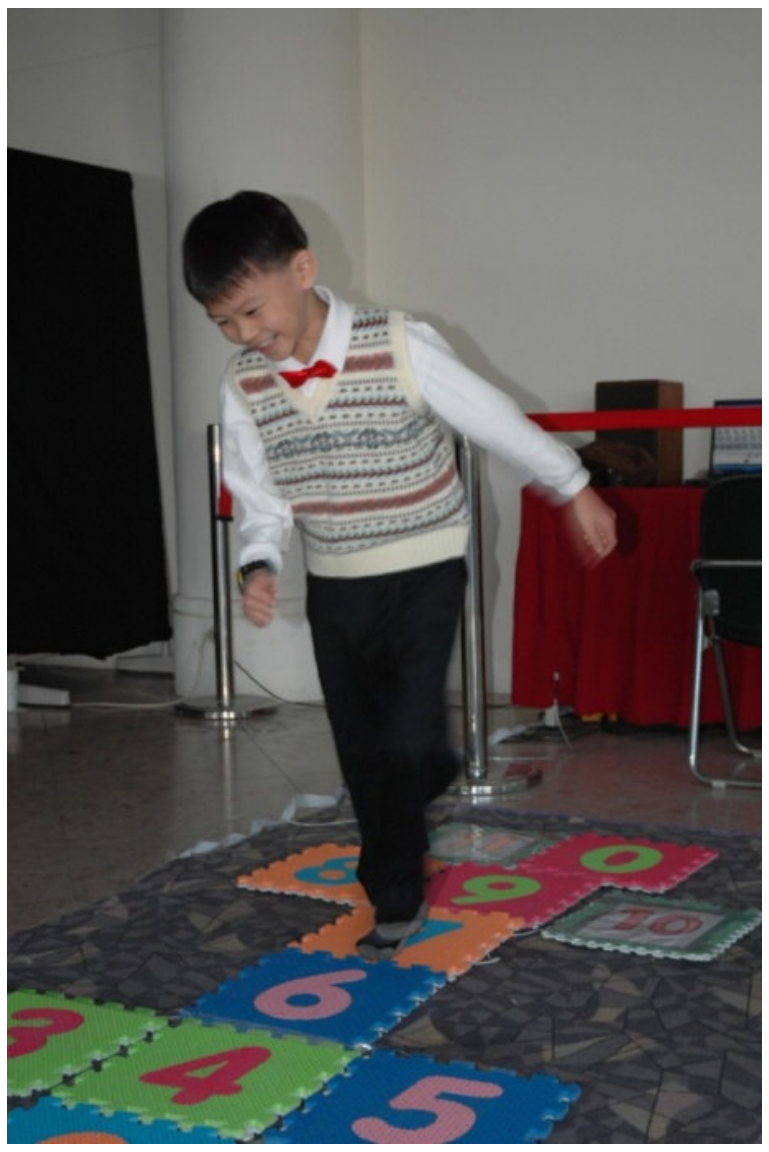

Figure 13. The New Type of Hopscotch Music Game Played

## 6. CONCLUSIONS

Although lots of the edutainment applications integrate video game into the education system, there are more and more innovative games using computer music for the sound system to achieve the purpose of music education. Instead of the video game based edutainment system, the new hopscotch is a set of hyperinstrument music game with automated ad collaborated composition functions to make the traditional hopscotch game more interesting. The computer music auxiliary game provides a variety of interests from the features of the game, especially the interaction between children and the music game, to generate many music elements involving pitch, rhythm, timbre, etc., to spontaneously produce the polyphonic effect while multi-user participators play the game. The generative music plays an important role to integrate with the kids' motion, to establish an interactive multimedia game, which well extends the usage of the traditional game. Some music parameters including rhythm variety, pitch sensation, texture change, sound response, dynamic variance, and timbre change have been well tested, and hopefully more music gestures can be presented by the system derived from the children's movement. Therefore the innovated hopscotch is a prototype of the interactive computer music game which can be extent into the fields of interactive multimedia game or installation art in the future.

## ACKNOWLEDGEMENTS

The authors appreciate the great help by Mr. Huang, T. L. in ITRI (Industrial Technology Research Institute of Taiwan, R.O.C), Prof. Yeh, J. H. in National Hsinchu University of Education, and the support from National Science Council of Taiwan: NSC98-2221-E-155-073.

**Authors**

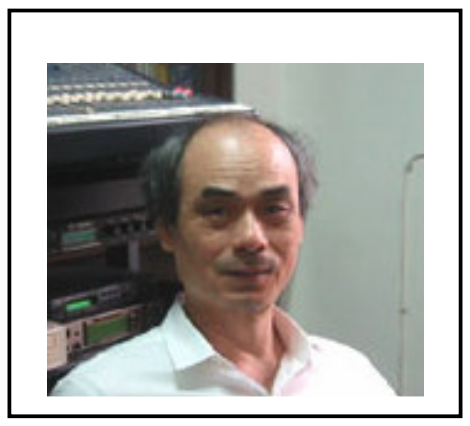

**Shing-Kwei Tzeng**, 1981 graduated with "Prüfung der künstlerlischen Reifer" at the Musikhochschule in Freiburg. 1981~2005 he is the Professor at the Music Department of National Taiwan Normal University. 1986~87 he won Scholarship of the French Government, studied film music at Ecole Normal de Musique de Paris/France and graduated with Diploma. 1987 he did the research at IRCAM, Paris. 1999 he founded Taiwan Computer Music Society and was selected as its first president until 2002 (re-elected in 2005–07). 2002~03 he won Fulbright Scholar as visiting scholar at CCRMA, Stanford, and became the guest composer at the College of Music, University of North Texas. 2005 he took up a professorship at the Department of Information Communications, Kainan University. He is a member of ACL (Asian Composers' League), ISCM (International Society for Contemporary Music), and Society of Taiwanese Music Education. He has won Le Premiere Grand Prix de 3eme Concour International Composition pour Orque 1986 ST. Remy/France, and Le Premiere Prix de Ville D'Avry International Library for Contemporary Music 1984 Paris/France.

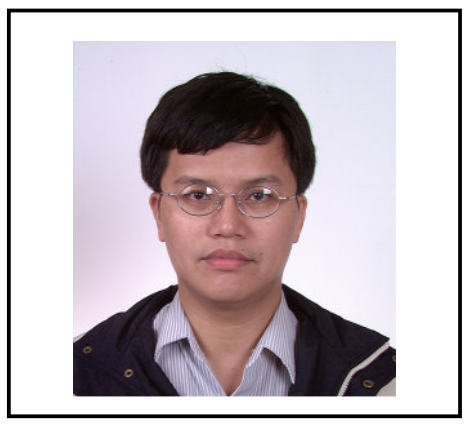

**Chih-Fang Huang**, the Assistant Professor at the Department of Information Communication at Yuan Ze University, was born in 1965 in Taipei city, Taiwan. He is the also the chairman of TCMA (Taiwan Computer Music Association) currently. He acquired both a PhD in mechanical engineering and a master's degree in music composition in 2001 and 2003 respectively from National Chiao Tung University. His research papers include many fields, such as automated music composition, virtual reality, and the intermedia integration.